\documentclass[aps,prl,reprint,               amsmath,amssymb,longbibliography]{revtex4-2}
\usepackage[T1]{fontenc}
\usepackage{lmodern}
\usepackage{graphicx}
\usepackage{booktabs}
\usepackage[hidelinks]{hyperref}
\newcommand{\Z}{\mathbb Z}
\newcommand{\Q}{\mathbb Q}
\newcommand{\C}{\mathbb C}
\newcommand{\dd}{\mathrm d}

\newcommand{\heading}[1]{\par\medskip\noindent\textit{#1.---}}
\makeatletter
\AtBeginDocument{\immediate\write\@auxout{\string\citation{isingControl}}}
\makeatother

\begin{document}
\title{Droplets obstruct Onsager-type solutions of the Ising model in a magnetic field}

\author{Ran Luo$^2$}
\author{Jiahua Tian$^1$}

\affiliation{%
 $^1$School of Physics, East China Normal University, Shanghai, China, 200241
}
\affiliation{%
 $^2$School of Physics, Peking University, Beijing, China, 100871
}

\begin{abstract}
Onsager solved the two-dimensional Ising model in zero field, but no solution is known in a magnetic field. We prove that in any dimension $d\ge2$, at every fixed dimensionless field outside a countable set, the free energy obeys no algebraic differential equation in the coupling, even locally. The obstruction arises from droplet geometry: the maximal droplet volume grows as the $d/(d-1)$ power of its boundary size, while an integer band in the pressure coefficients protects the extremal contributions against cancellation. Our result excludes Onsager-type solutions and closed forms in elliptic, hypergeometric, or Painlev\'e functions of the coupling at these fields.
\end{abstract}
\maketitle

\heading{Introduction}
The zero-field square-lattice Ising model was solved by Onsager~\cite{Onsager1944}, and its spontaneous magnetization was derived by Yang~\cite{Yang1952}. In a field, the scaling limit at the critical temperature is integrable and described by the $E_8$ field theory~\cite{Zamolodchikov1989,Bazhanov1994}. On the lattice, the partition function in a field can be written as a sum of Pfaffians~\cite{Chertkov2008,Gomez2009}, and the low-temperature expansion of the free energy is organized by droplets, connected islands of reversed spins. In series that keep the full field dependence at each order, the leading terms come from square and rectangular droplets~\cite{Meissner2023}. The lattice free energy in a field is nevertheless unknown, and studies of series and of partition-function zeros suggest that the model in a field is not integrable~\cite{McCoyMaillard2012,Assis2017}. In many exactly solved models, thermodynamic functions obey differential equations~\cite{Onsager1944,Kasteleyn1961,FanWu1970,McCoyWu1973,WuMcCoyTracyBarouch1976,Baxter1980,Joyce1988}, and series expansions test whether a lattice model has a solution of this kind~\cite{GuttmannEnting1996}. For self-avoiding polygons, such a test led to a proof that the generating function with separate weights for horizontal and vertical steps obeys no linear differential equation~\cite{Rechnitzer2003,Rechnitzer2006}. Whether the zero-field Ising susceptibility obeys an algebraic differential equation remains open~\cite{Guttmann2016}.

Onsager's free energy~\cite{Onsager1944}, like those of dimer models~\cite{Kasteleyn1961} and free-fermion vertex models~\cite{FanWu1970}, is a Brillouin-zone integral of the logarithm of a trigonometric polynomial. With the dimensionless coupling $K=J/(k_BT)$ and field $H=h/(k_BT)$, we call $p$ an Onsager-type pressure on an interval of couplings when it has the form
\begin{equation}
	p=R+\sum_jc_j\int_{[-\pi,\pi]^k}\frac{\dd^k\theta}{(2\pi)^k}\log Q_j\bigl(e^{i\theta_1},\ldots,e^{i\theta_k}\bigr),
	\label{eq:FF}
\end{equation}
on that interval with finitely many terms, constants $c_j$, a function $R$ algebraic in the Boltzmann weights $e^{\pm K}$ and $e^{\pm H}$, and trigonometric polynomials $Q_j$ whose coefficients are algebraic in these weights and which have no zeros at real momenta, as for gapped free fermions, so that each logarithm stays finite. The derivative of such a pressure in $K$ obeys a linear differential equation in $u=e^{-4K}$, which in Onsager's case is solved by a Gauss hypergeometric function~\cite{DLMF}, so the pressure itself obeys an algebraic differential equation~\cite{Lipshitz1988,Zeilberger1990,SM3}. Beyond free fermions, a solution in a field may use elliptic functions or the Painlev\'e transcendents that describe zero-field correlations~\cite{WuMcCoyTracyBarouch1976,DLMF}. All of these candidates obey \emph{algebraic differential equations} in the coupling. We call a function $\Phi(u)$ \emph{differentially algebraic} when
\begin{equation}
	\mathcal P\bigl(u,\Phi,\ldots,\partial_u^s\Phi\bigr)=\sum_{\mu,\{\nu_j\}}c_{\mu,\{\nu_j\}}\,u^{\mu}\prod_{j=0}^{s}(\partial_u^j\Phi)^{\nu_j}=0
	\label{eq:DA}
\end{equation}
for some finite order $s$, where the sum runs over finitely many nonnegative integer exponents $\mu$ and $\{\nu_j\}$ with complex coefficients $c_{\mu,\{\nu_j\}}$, not all zero, and \emph{differentially transcendental}, or hypertranscendental, otherwise~\cite{Rubel1989,Manssour2024}, with the Gamma function as a classic example~\cite{Holder1886}. We denote the $N$ monomials in~\eqref{eq:DA} by $\mathcal M_i=u^{\mu_i}\prod_{j=0}^{s}(\partial_u^j\Phi)^{\nu_{ij}}$, $i=1,\ldots,N$. Differentially algebraic functions are closed under arithmetic operations and composition~\cite{Manssour2024}, so differential algebraicity in $u$ and in $K$ are equivalent.

We prove that the free energy of the Ising model in a magnetic field is differentially transcendental in the coupling $K$ at every fixed dimensionless field $H$ outside a countable set, in every dimension $d\ge2$. Solutions of Onsager's type and closed forms in elliptic, hypergeometric, or Painlev\'e functions of the coupling are therefore impossible at almost every field. On the square lattice, an $r\times r$ droplet contributes $u^{2r}z^{r^2}$, where $z=e^{-2H}$ is the fugacity. Combinations of smaller droplets with the same total boundary contain at least $2r-2$ fewer reversed spins, so only single droplets reach the highest powers of $z$, which have integer coefficients. We show that a factor $B^{r^2}$ is protected by this integer band in the reduced denominator of the coefficient of $u^{2r}$ at a rational fugacity $z=A/B\in(0,1)$, with coprime $A$ and $B$, for every $r\ge3$. At $n=2r$ the factor $B^{r^2}$ grows faster than the bound $e^{Cn(\log n)^2}$ on the denominators of power series with rational coefficients that obey algebraic differential equations~\cite{Polya1916,Popken1935,SibuyaSperber1981,KrattenthalerRivoal2025}.

\heading{Model and result}
We consider the Ising model on the square lattice $\Z^2$, with spins $\sigma_i=\pm1$ and Hamiltonian
\begin{equation}
	\mathcal H=-J\sum_{\langle ij\rangle}\sigma_i\sigma_j-h\sum_i\sigma_i,\qquad J>0,\quad h>0,
	\label{eq:Hamiltonian}
\end{equation}
where $J$ is the exchange coupling, $h$ the magnetic field, and $\langle ij\rangle$ runs over the nearest-neighbor bonds. Here $u$ is the weight of two broken bonds, since each broken bond between antiparallel spins costs $2J$ more than a parallel one, and the fugacity $z$ is the weight of one reversed spin, since each spin reversed against the field costs $2h$, so $0<z<1$ for positive $h$. The dimensionless pressure and the free energy per site are
\begin{equation}
	p(K,H)=\lim_{N\to\infty}\frac{1}{N^2}\log Z_N,\qquad f=-k_BT p,
\end{equation}
where $Z_N$ is the partition function of an $N\times N$ box. The limit is independent of the boundary conditions, because changing the boundary spins alters only the $4N$ boundary bonds, by at most $2J$ each~\cite{Griffiths1964,FriedliVelenik2017}. Spin reversal gives $p(K,-H)=p(K,H)$, so we take $h>0$. We consider the classical low-temperature expansion of the pressure~\cite{SykesEssamGaunt1965},
\begin{equation}
	p=2K+H+\Phi(u,z),\quad\Phi(u,z)=\sum_{m\ge2}L_m(z)u^m
	\label{eq:series}
\end{equation}
where $2K+H$ is the pressure of the fully aligned ground state for positive $h$, and $\Phi$ is the contribution of the droplets, which starts at $u^2$, the weight of the four broken bonds around a single reversed spin. Our argument concerns the coefficients $L_m$, and it constrains the physical pressure because for every $|z|<1$ the series converges near $u=0$ to the physical $\Phi$. We show this convergence with the Lee--Yang circle theorem, which bounds the coefficients of the expansion in $z$ uniformly at every temperature, and with the fact that these coefficients are polynomials in $u$ of bounded degree~\cite{LeeYang1952,SM1}.

Our main result for the square lattice is that, for every rational $0<z<1$, $\Phi(u,z)$ is differentially transcendental in $u$, and so is $p=-\tfrac12\log u+H+\Phi$, since $-\tfrac12\log u+H$ is differentially algebraic in $u$. The magnetization at the same fixed $z$,
\begin{equation}
	M=\partial_Hp=1-2z\partial_z\Phi,
	\label{eq:observables}
\end{equation}
is differentially transcendental as well. The result extends to every $0<z<1$ that is not an algebraic integer, including every transcendental $z$~\cite{SM2}. By the Lindemann--Weierstrass theorem~\cite{Baker1975}, $z$ is transcendental whenever $H$ is a nonzero algebraic number, so the result holds at every positive algebraic value of $H$. It remains open whether the free energy is differentially algebraic at fugacities that are algebraic integers, which form a countable set.

\heading{Area gap}
Let $S$ be a nonempty finite set of reversed spins with $n(S)=|S|$ sites, and let $|\partial S|$ count the bonds connecting $S$ to its complement. By~\eqref{eq:Hamiltonian}, each broken bond costs $2J$ and each reversed spin $2h$ relative to the fully aligned ground state~\cite{Enting1978,Meissner2023}, so
\begin{equation}
	\begin{aligned}
		\Delta E(S)&=2J|\partial S|+2h|S|,\\
		e^{-\Delta E/(k_BT)}&=u^{m(S)}z^{n(S)},\qquad m(S)=\frac{|\partial S|}{2},
	\end{aligned}
	\label{eq:droplet}
\end{equation}
which is an integer because $|\partial S|=4|S|-2e(S)$, where $e(S)$ counts the bonds joining two sites of $S$. Hence a square droplet of side $r$ costs $8Jr$ in exchange energy and $2hr^2$ in field energy and contributes $u^{2r}z^{r^2}$.

\begin{figure*}[t]
\centering
\setlength{\unitlength}{1pt}
\begin{picture}(410,112)
 \put(0,95){\makebox(136,14){\small (a) Bounding box}}
 \put(137,95){\makebox(136,14){\small (b) One square droplet}}
 \put(274,95){\makebox(136,14){\small (c) Two droplets}}
 \put(36,53){\line(1,0){16}}
 \put(36,69){\line(1,0){16}}
 \put(52,37){\line(1,0){16}}
 \put(52,53){\line(1,0){16}}
 \put(52,69){\line(1,0){16}}
 \put(52,85){\line(1,0){16}}
 \put(68,37){\line(1,0){16}}
 \put(68,53){\line(1,0){16}}
 \put(68,69){\line(1,0){16}}
 \put(84,37){\line(1,0){16}}
 \put(84,53){\line(1,0){16}}
 \put(36,53){\line(0,1){16}}
 \put(52,37){\line(0,1){16}}
 \put(52,53){\line(0,1){16}}
 \put(52,69){\line(0,1){16}}
 \put(68,37){\line(0,1){16}}
 \put(68,53){\line(0,1){16}}
 \put(68,69){\line(0,1){16}}
 \put(84,37){\line(0,1){16}}
 \put(84,53){\line(0,1){16}}
 \put(100,37){\line(0,1){16}}
 \put(34,35){\dashbox{2}(68,52){}}
 \put(36,23){\makebox(64,10){\small $w_1=4$}}
 \put(0,56){\makebox(31,10)[r]{\small $w_2=3$}}
 \put(0,9){\makebox(136,13){\small $m=7,\quad n=7$}}
 \put(173,29){\framebox(64,64){}}
 \multiput(189,29)(16,0){3}{\line(0,1){64}}
 \multiput(173,45)(0,16){3}{\line(1,0){64}}
 \put(137,9){\makebox(136,13){\small $m=8,\quad n=16,\quad u^8z^{16}$}}
 \put(296,45){\framebox(48,48){}}
 \multiput(312,45)(16,0){2}{\line(0,1){48}}
 \multiput(296,61)(0,16){2}{\line(1,0){48}}
 \put(362,61){\framebox(16,16){}}
 \put(274,9){\makebox(136,13){\small $m=6+2=8,\quad n=9+1=10$}}
 \put(0,-2){\makebox(410,12){\small Each cell is a reversed spin, and the exterior is aligned.}}
\end{picture}
\caption{(a) A droplet occupying $w_1$ columns and $w_2$ rows fits in a $w_1\times w_2$ bounding box (dashed). Each occupied row and column crosses its boundary at least twice, so $m\ge w_1+w_2$. (b), (c) At $m=8$, a single square contains $16$ reversed spins, whereas several droplets contain at most $10$. This gap protects the highest-degree coefficients of $L_m(z)$. For $0<z<1$ the larger area of the square means stronger field suppression, so the square enters our argument through the small weight $z^{16}$.}
\label{fig:droplet}
\end{figure*}
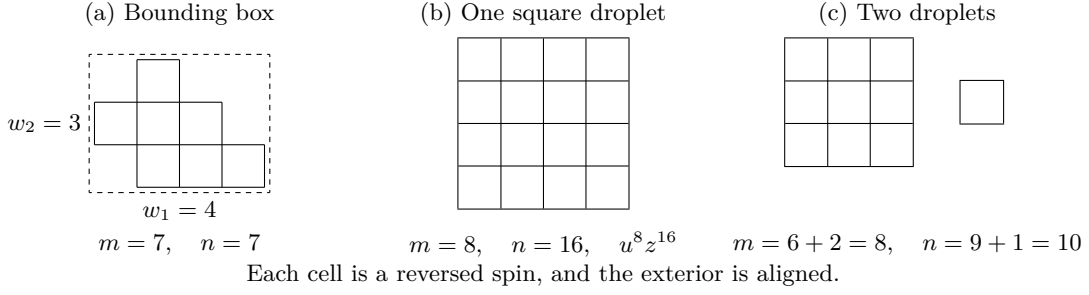

For any $S$ that occupies $w_1$ columns and $w_2$ rows [Fig.~\ref{fig:droplet}(a)], each occupied row and each occupied column contributes at least two boundary bonds. Hence $m\ge w_1+w_2$ and
\begin{equation}
	n\le w_1w_2\le A_m:=\left\lfloor\frac{m^2}{4}\right\rfloor
	\label{eq:isoperimetry}
\end{equation}
and equality is achieved only for a filled rectangle with $|w_1-w_2|\le1$. For $m=2r$ this is the $r\times r$ square and for $m=2r+1$ the $r\times(r+1)$ rectangle in its two orientations, in both cases up to translation.

For $k\ge2$ droplets $S_j$ with areas $n_j=|S_j|$ and half-perimeters $m_j\ge2$ summing to $m\ge4$, the bound~\eqref{eq:isoperimetry} for each droplet gives
\begin{equation}
	\sum_{j=1}^{k}n_j\le\sum_{j=1}^{k}A_{m_j}\le B_m:=A_{m-2}+1=A_m-(m-2),
	\label{eq:gap}
\end{equation}
since $A_m$ is convex in $m$, where $B_m$ is achieved by putting half-perimeter $m-2$ into one droplet and a single reversed spin into the other~\cite{SM1} [Fig.~\ref{fig:droplet}(c)]. By~\eqref{eq:gap}, several droplets with total half-perimeter $m$ contain at least $m-2$ fewer reversed spins than the most a single droplet can contain.

\heading{Integer band}
Since the pressure is independent of the boundary conditions~\cite{Griffiths1964,FriedliVelenik2017}, we compute $\Phi$ in an $a\times b$ box $\Lambda_{a,b}$ surrounded by spins fixed to $+1$. Dividing the partition function by the weight of the all-plus state~\cite{Enting1978}, we obtain
\begin{equation}
	\begin{aligned}
		P_{a,b}(u,z)&=\sum_{S\subseteq\Lambda_{a,b}}u^{m(S)}z^{|S|}\in\Z[u,z],\\
		P_{a,b}(0,z)&=1,
	\end{aligned}
	\label{eq:finiteP}
\end{equation}
and $\Phi=\lim_{a,b\to\infty}(ab)^{-1}\log P_{a,b}$ order by order in $u$~\cite{SM1}.

The droplets of a configuration neither overlap nor touch, so $P_{a,b}$ is the partition function of a gas of droplets with hard-core exclusion~\cite{GruberKunz1971}, and $\log P_{a,b}$ is the Mayer cluster expansion of this gas~\cite{Mayer1937,Brydges1986,Bissacot2010,SM1}, which is a sum over Mayer clusters. A Mayer cluster is a list of droplets, repeats included, joined into one connected group by overlaps (shared sites) and contacts (neighboring sites), and it carries the product of the weights of its droplets $S_j$ times a combinatorial factor, so its powers of $u$ and $z$ are $\sum_jm(S_j)$ and $\sum_jn(S_j)$, respectively, and a single droplet is a Mayer cluster with combinatorial factor one.

At order $u^m$ there are finitely many Mayer clusters up to translation, each with width plus height at most $m$~\cite{SM1}. A Mayer cluster of size $w_1\times w_2$ fits into the box in $(a-w_1+1)(b-w_2+1)$ positions. Hence $[u^m]\log P_{a,b}=L_m\,ab+\lambda_ma+\lambda'_mb+\kappa_m$ for $a,b\ge m$, where $[x^n]F$ denotes the coefficient of $x^n$ in $F$, since only the bulk term survives in $\Phi=\lim_{a,b\to\infty}(ab)^{-1}\log P_{a,b}$. Taking the mixed difference $\Delta_a\Delta_b$, with $\Delta_aF_{a,b}=F_{a,b}-F_{a-1,b}$, we remove the boundary and corner terms without dividing by the volume,
\begin{equation}
	L_m=[u^m]\Delta_a\Delta_b\log P_{a,b},\qquad a,b\ge m+1,
	\label{eq:bulkdiff}
\end{equation}
so $L_m$ is an integer combination of finite-box coefficients.

\begin{figure}[t]
\setlength{\unitlength}{1pt}
\begin{picture}(246,54)
 \put(132,8){\dashbox{2}(108,46){}}
 \put(6,8){\line(1,0){234}}
 \put(15,8){\line(0,1){2}}
 \put(15,0){\makebox(0,6){\footnotesize 4}}
 \put(15,27){\makebox(0,0){\footnotesize $-\frac{209}{4}$}}
 \put(33,8){\line(0,1){2}}
 \put(33,0){\makebox(0,6){\footnotesize 5}}
 \put(33,27){\makebox(0,0){\rotatebox{90}{\footnotesize $926$}}}
 \put(51,8){\line(0,1){2}}
 \put(51,0){\makebox(0,6){\footnotesize 6}}
 \put(51,27){\makebox(0,0){\rotatebox{90}{\footnotesize $-1651$}}}
 \put(69,8){\line(0,1){2}}
 \put(69,0){\makebox(0,6){\footnotesize 7}}
 \put(69,27){\makebox(0,0){\rotatebox{90}{\footnotesize $-486$}}}
 \put(87,8){\line(0,1){2}}
 \put(87,0){\makebox(0,6){\footnotesize 8}}
 \put(87,27){\makebox(0,0){\footnotesize $\frac{389}{2}$}}
 \put(105,8){\line(0,1){2}}
 \put(105,0){\makebox(0,6){\footnotesize 9}}
 \put(105,27){\makebox(0,0){\rotatebox{90}{\footnotesize $540$}}}
 \put(123,8){\line(0,1){2}}
 \put(123,0){\makebox(0,6){\footnotesize 10}}
 \put(123,27){\makebox(0,0){\rotatebox{90}{\footnotesize $461$}}}
 \put(141,8){\line(0,1){2}}
 \put(141,0){\makebox(0,6){\footnotesize 11}}
 \put(141,27){\makebox(0,0){\rotatebox{90}{\footnotesize $310$}}}
 \put(159,8){\line(0,1){2}}
 \put(159,0){\makebox(0,6){\footnotesize 12}}
 \put(159,27){\makebox(0,0){\rotatebox{90}{\footnotesize $151$}}}
 \put(177,8){\line(0,1){2}}
 \put(177,0){\makebox(0,6){\footnotesize 13}}
 \put(177,27){\makebox(0,0){\footnotesize $68$}}
 \put(195,8){\line(0,1){2}}
 \put(195,0){\makebox(0,6){\footnotesize 14}}
 \put(195,27){\makebox(0,0){\footnotesize $22$}}
 \put(213,8){\line(0,1){2}}
 \put(213,0){\makebox(0,6){\footnotesize 15}}
 \put(213,27){\makebox(0,0){\footnotesize $6$}}
 \put(231,8){\line(0,1){2}}
 \put(231,0){\makebox(0,6){\footnotesize 16}}
 \put(231,27){\makebox(0,0){\footnotesize $1$}}
 \put(132,44){\makebox(108,8){\footnotesize integer band}}
 \put(6,44){\makebox(126,8){\footnotesize one or more droplets}}
\end{picture}
\caption{Coefficients $[z^n]L_8$ of~\eqref{eq:series}~\cite{Sykes1973}. In the integer band, only single droplets contribute and the coefficients count droplets up to translation, ending with the $4\times4$ square at $n=16$. Fractions appear only below the band, with denominators dividing $m=8$.}
\label{fig:band}
\end{figure}
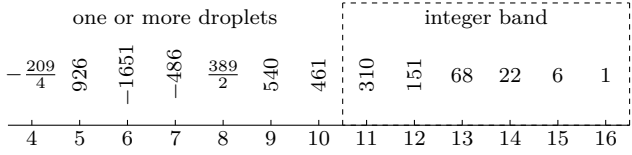

By~\eqref{eq:gap}, a Mayer cluster of more than one droplet at order $u^m$ has a power of $z$ at most $B_m$, so only single droplets contribute to the powers $z^n$ with $B_m<n\le A_m$, each with coefficient one per translate. We call this range of width $m-2$ the \emph{integer band} of $L_m$ (Fig.~\ref{fig:band}). In this band, $[z^n]L_m$ is the number of droplets $S$ with $m(S)=m$ and $n(S)=n$ up to translation~\cite{SM1}, and by~\eqref{eq:isoperimetry}
\begin{equation}
 \deg_zL_m=A_m,\quad
 [z^{r^2}]L_{2r}=1,\quad
 [z^{r(r+1)}]L_{2r+1}=2.
 \label{eq:top}
\end{equation}
Below the band the combinatorial factors produce fractions, and we use the logarithmic derivative to show that their denominators divide $m$. Since $P_{a,b}\in\Z[u,z]$ and $P_{a,b}(0,z)=1$ by~\eqref{eq:finiteP}, $1/P_{a,b}=\sum_{k\ge0}(1-P_{a,b})^k$ is a series in $u$ with coefficients in $\Z[z]$, and so is $u\partial_u\log P_{a,b}=u\,\partial_uP_{a,b}/P_{a,b}$, whose coefficient of $u^m$ is $m[u^m]\log P_{a,b}\in\mathbb Z[z]$. With~\eqref{eq:bulkdiff} and the integer band we obtain
\begin{equation}
	mL_m\in\Z[z],\qquad [z^n]L_m\in\Z\quad\hbox{for }n>B_m.
	\label{eq:integrality}
\end{equation}

\heading{Denominator growth}
We focus on even orders $m=2r\ge4$. We separate the integer-band part $S_{2r}(z)=\sum_{n>B_{2r}}([z^n]L_{2r})\,z^n$ of $L_{2r}$ and write
\begin{equation}
	L_{2r}(z)=S_{2r}(z)+\frac{R_{2r}(z)}{2r}.
	\label{eq:split}
\end{equation}
By~\eqref{eq:top} and~\eqref{eq:integrality}, $S_{2r}\in\Z[z]$ is monic of degree $r^2$, and $R_{2r}=2rL_{2r}-2rS_{2r}\in\Z[z]$. By construction, $R_{2r}$ contains only the powers $z^n$ with $n\le B_{2r}=r^2-2r+2$. We now fix $z=A/B$ and a prime $\ell$ dividing $B$. For a nonzero rational number $x$ we write $x=\ell^{v_\ell(x)}x'$, where the numerator and denominator of $x'$ are not divisible by $\ell$, and put $a_\ell=v_\ell(B)>0$. The square droplet contributes
\begin{equation}
	v_\ell\bigl((A/B)^{r^2}\bigr)=-a_\ell r^2
\end{equation}
since $v_\ell(xy)=v_\ell(x)+v_\ell(y)$~\cite{SM2}. Every other term of the integer polynomial $S_{2r}(A/B)$ has $v_\ell\ge-a_\ell(r^2-1)$, and the lower-degree part obeys
\begin{equation}
	v_\ell\Bigl(\frac{R_{2r}(A/B)}{2r}\Bigr)\ge -a_\ell(r^2-2r+2)-v_\ell(2r).
	\label{eq:remainder}
\end{equation}
The integer band of $L_{2r}$ has width $2r-2$, whereas dividing by $2r$ adds at most $v_\ell(2r)$ powers of $\ell$ to the denominator, and $v_\ell(2r)\le\log_\ell(2r)\le\log_2(2r)$ because $\ell^{v_\ell(2r)}$ divides $2r$. Since $a_\ell\ge1$, we have $a_\ell(2r-2)\ge2r-2>\log_2(2r)\ge v_\ell(2r)$ for every $r\ge3$, so by~\eqref{eq:remainder} the square droplet is the only term with the most negative exponent of $\ell$. By $v_\ell(x+y)\ge\min\{v_\ell(x),v_\ell(y)\}$~\cite{SM2}, the sum $L_{2r}(A/B)-(A/B)^{r^2}$ of all other terms also has $v_\ell>-a_\ell r^2$. Because $v_\ell(x+y)=\min\{v_\ell(x),v_\ell(y)\}$ when $v_\ell(x)\ne v_\ell(y)$, adding the square term gives
\begin{equation}
	v_\ell\bigl(L_{2r}(A/B)\bigr)=-v_\ell(B)\,r^2 \quad\hbox{for every }r\ge3.
	\label{eq:valuation}
\end{equation}
Since $L_{2r}(A/B)\,d_{2r}$ is an integer, where $d_{2r}$ is the reduced denominator of $L_{2r}(A/B)$, we have $v_\ell(d_{2r})\ge-v_\ell\bigl(L_{2r}(A/B)\bigr)=v_\ell(B)\,r^2$ for every prime $\ell$ dividing $B$. Hence $d_{2r}$ is divisible by $\prod_{\ell\mid B}\ell^{v_\ell(B)r^2}=B^{r^2}$. In the other direction, $2rL_{2r}\in\Z[z]$ has degree $r^2$ by~\eqref{eq:top} and~\eqref{eq:integrality}, and $B^{r^2}z^n=A^nB^{r^2-n}$ is an integer at $z=A/B$ for every $n\le r^2$, so $2rB^{r^2}L_{2r}(A/B)\in\Z$ and $d_{2r}$ divides $2rB^{r^2}$. Together we have $B^{r^2}\mid d_{2r}\mid 2rB^{r^2}$, hence
\begin{equation}
	\log d_{2r}=r^2\log B+O(\log r).
	\label{eq:denomgrowth}
\end{equation}
Consequently, the area of the square droplet determines the growth of the denominators up to a logarithmic correction.

\heading{No algebraic differential equation}
A differentially algebraic power series with algebraic coefficients $g_n$ has denominators $d_n$, the least positive integers for which $d_ng_n$ are algebraic integers, obeying
\begin{equation}
	\log d_n=O\!\left(n[\log(n+2)]^2\right).
	\label{eq:polya}
\end{equation}
This bound goes back to P\'olya and Popken~\cite{Polya1916,Popken1935}. It is proved in~\cite{Mahler1976} and stated in~\cite{KrattenthalerRivoal2025} for every equation~\eqref{eq:DA}, since an equation with complex coefficients can be replaced by one with algebraic coefficients~\cite{Mahler1976,Kolchin1973,SM2}. Bounds of this type have been used before to show that a given power series is not differentially algebraic~\cite{Katriel2003}. Because $r(\log r)^2$ grows more slowly than $r^2$,~\eqref{eq:denomgrowth} violates~\eqref{eq:polya} at the orders $n=2r$ for all large $r$. We have thus proved the main result at rational fugacity. 

At an algebraic $z$ that is not an algebraic integer, some valuation $v$ of $\Q(z)$ that extends $v_\ell$ for a prime $\ell$ has $v(z)<0$~\cite{Neukirch1999}. It obeys the same rules as $v_\ell$~\cite{SM2}, so the comparison leading to~\eqref{eq:valuation} again isolates the square and gives $v(L_{2r}(z))=v(z)\,r^2$ for sufficiently large $r$. Since $d_{2r}L_{2r}(z)$ is an algebraic integer and $v$ agrees with $v_\ell$ on ordinary integers, we have $v_\ell(d_{2r})\ge -v(z)r^2$. Consequently, $\log d_{2r}\ge -v(z)r^2\log\ell$, contradicting~\eqref{eq:polya}. 

For transcendental $z$ we first keep $z$ as a variable. We expand each monomial as $\mathcal M_i=\sum_{n\ge0}t_{ni}(z)\,u^n$. By~\eqref{eq:integrality}, $L_m\in\Q[z]$, hence by~\eqref{eq:series} every $t_{ni}$ lies in $\Q[z]$. At each order of $u$,~\eqref{eq:DA} then becomes a linear equation for the coefficients $c_i=c_{\mu_i,\{\nu_{ij}\}}$ of the $\mathcal M_i$,
\begin{equation}
	\sum_{i=1}^{N}t_{ni}(z)\,c_i=0,\qquad n\ge0.
	\label{eq:linear}
\end{equation}
By the rational case, at any rational $z_0\in(0,1)$ these equations have only the solution $c_i=0$, so some $N\times N$ minor $D(z)\in\Q[z]$ of the matrix $(t_{ni}(z))_{n\ge0,\,1\le i\le N}$ of~\eqref{eq:linear} obeys $D(z_0)\ne0$. Thus $D$ is a nonzero polynomial with rational coefficients, which cannot vanish at any transcendental $z$, so there the $N$ equations of~\eqref{eq:linear} that form $D$ force $c_i=0$~\cite{SM2}. Since the coefficients of~\eqref{eq:DA} are not all zero, $\Phi(u,z)$ fails to obey~\eqref{eq:DA} at transcendental $z$. Combining this with the rational and algebraic cases, we find that $\Phi(u,z)$ is differentially transcendental in $u$ at every field outside the countable set of fields whose fugacity $z$ is an algebraic integer.

\heading{No Onsager-type solution}
We now use our result to constrain exact solutions. Suppose that at fixed field $H>0$ a closed-form expression reproduces the pressure on an interval of $K$ and obeys an algebraic differential equation $\mathcal P(K,p,\partial_Kp,\ldots,\partial_K^sp)=0$ of the form~\eqref{eq:DA} there. At fixed $H>0$ the pressure is real analytic in $K$ for all $K>0$~\cite{Ott2020}, so by the identity theorem it obeys this equation for all $K>0$, in particular at small $u$, where $\Phi=p-2K-H$ is given by the convergent series~\eqref{eq:series}. Since differential algebraicity in $u$ and in $K$ are equivalent, $\Phi$ would obey an equation of the form~\eqref{eq:DA} near $u=0$, which we have excluded at every field outside the countable set~\cite{SM3}.

Every solution of Onsager's type~\eqref{eq:FF} obeys such an equation, since its $u$-derivative obeys a linear Picard--Fuchs equation~\cite{Lipshitz1988,Zeilberger1990,SM3}. By our result it differs from the Ising pressure on every interval of the coupling, at every field outside the countable set. Closed forms built from elliptic, hypergeometric, or Painlev\'e  functions with arguments differentially algebraic in $K$ with fixed parameters fail in the same way, because these functions obey algebraic differential equations in their arguments~\cite{SM3}. In particular, no such expression agrees with the pressure on any open set of couplings and fields.

\heading{Every dimension}
On the hypercubic lattice $\Z^d$, the Hamiltonian~\eqref{eq:Hamiltonian} gives $p=dK+H+\Phi(u,z)$~\cite{ButeraPernici2016}, a set $S$ of reversed spins again contributes $u^{m(S)}z^{|S|}$ with $m(S)=|\partial S|/2$, and a single reversed spin costs $u^d$. From the Loomis--Whitney inequality~\cite{LoomisWhitney1949} we obtain
\begin{equation}
	|\partial S|\ge2d\,|S|^{(d-1)/d},
	\label{eq:LW}
\end{equation}
with equality only for a cube~\cite{EllisFriedgutKindlerYehudayoff2016,SM4}. A cube of side $r$ has $m=dr^{d-1}$ and contributes $u^{dr^{d-1}}z^{r^d}$, so its volume grows with $m$ as $r^d=(m/d)^{d/(d-1)}$. A Mayer cluster of several droplets with the same total boundary has at least $\tfrac{d}{d-1}(r-1)-1$ fewer reversed spins than the cube, so only single droplets contribute to the powers of $z$ above this bound, each with coefficient one per translate. In this integer band of $L_{dr^{d-1}}$ the cube is the only droplet of volume $r^d$, so $[z^{r^d}]L_{dr^{d-1}}=1$. At $z=A/B$ the cube term then makes the reduced denominator $d_m$ of $L_m(A/B)$ divisible by $B^{r^d}$ for large $r$, as in two dimensions, so $\log d_m\ge(m/d)^{d/(d-1)}\log B$ grows faster than $m[\log(m+2)]^2$ and violates~\eqref{eq:polya}. Convergence of the low-temperature series near $u=0$, established with the Lee--Yang theorem as in two dimensions~\cite{LeeYang1952}, and analyticity in $K$ at fixed $H\ne0$~\cite{Ott2020} also hold on $\Z^d$, so our main result and its consequences for exact solutions hold for every $d\ge2$, including the three-dimensional Ising model in a field~\cite{SM4}.

\heading{Discussion}
The droplet theory of condensation rests on the same competition between volume and boundary as our argument, and it was predicted in~\cite{Andreev1964,Fisher1967,Langer1967} that droplets give the free energy an essential singularity at $H=0$ below the critical temperature. For the low-temperature Ising model it was proved in~\cite{Isakov1984} that the $k$th field derivative of the free energy at $H=0^+$ grows like $(k!)^{d/(d-1)}$ in $d$ dimensions, and this result was extended to general lattice models with two ground states in~\cite{FriedliPfister2004}. The exponent $d/(d-1)$ of this growth is the inverse of the exponent in~\eqref{eq:LW}, and it sets the volume $r^d=(m/d)^{d/(d-1)}$ of the cube droplets. A single geometric exponent thus controls both the essential singularity at zero field and the obstruction to an algebraic differential equation at nonzero field. The obstruction lies in the denominators alone, while the coefficients themselves grow at most exponentially, as the analyticity of $\Phi$ near $u=0$ requires~\cite{SM1}.

At zero field, $z=1$, the area weight disappears together with the denominator growth of~\eqref{eq:denomgrowth}, consistent with Onsager's solution, whose free energy obeys a linear differential equation in $u$. The model was also solved at the imaginary field $H=i\pi/2$, where $z=-1$~\cite{LeeYang1952}. Both solved fugacities, $z=1$ and $z=-1$, are algebraic integers, consistent with the absence of the obstruction there. In one dimension~\eqref{eq:LW} reduces to $|\partial S|\ge2$, and a reversed interval of any length has exactly two boundary bonds, so the boundary no longer bounds the volume. Consistently, the exact pressure in a field~\cite{Ising1925} is the logarithm of an algebraic function of $u$ and $z$, so it is differentially algebraic in $u$. Therefore, the obstruction needs both the field and a dimension $d\ge2$.

The thermodynamic limit is also essential. In any finite box, $P$ is a polynomial in $u$, so $y=\log P$ satisfies the first-order linear equation $Py'-P'=0$. On a strip of fixed width $w$, the pressure is $p_w=w^{-1}\log\lambda_w$, where $\lambda_w$ is the largest eigenvalue of a finite transfer matrix~\cite{KramersWannier1941,Montroll1941,Baxter1982,SM1}. Since $\lambda_w$ is algebraic in the Boltzmann weights, $p_w$ is differentially algebraic in the coupling~\cite{SM1}. There the square droplets cannot grow beyond the width, and the growth of~\eqref{eq:denomgrowth} is lost. Hence, only the planar limit, which brings in square droplets of every size, is differentially transcendental at fields outside the countable set.

Mean-field theory, by contrast, has a pressure $\bar p$ that obeys an equation of the form~\eqref{eq:DA}. For example, for the Curie--Weiss energy $-\tfrac{J}{2N}(\sum_i\sigma_i)^2-h\sum_i\sigma_i$, the magnetization $\bar M=\tanh(K\bar M+H)$ obeys $\dd \bar M/\dd K=\bar M(1-\bar M^2)/[1-K(1-\bar M^2)]$, and $\partial_K\bar p=\bar M^2/2$~\cite{FriedliVelenik2017,Baxter1982,SM3}. The same holds whenever finitely many order parameters are fixed by nondegenerate equations like $\bar M=\tanh(K\bar M+H)$ and the pressure is an elementary function of them, so no mean-field approximation of this kind reproduces the lattice pressure on an interval of $K$ at a field outside the countable set.

An exact solution of the Ising model in a magnetic field, if one exists, cannot admit a representation of Onsager's type~\eqref{eq:FF} or any other representation differentially algebraic in the coupling $K$, at any fixed fugacity $z\in(0,1)$ outside the countable set of algebraic integers. The generating function of the square droplets, $\theta(u)=\sum_{r\ge0}z^{r^2}u^{2r}$, is a partial theta function, and it obeys the $q$-difference equation $\theta(u)=1+zu^2\,\theta(zu)$~\cite{Kostov2019}, in which $z$ sets the rescaling $u\to zu$. Difference equations of this type and integral representations remain open routes to the free energy.

\heading{Data availability} The data used to generate Fig.~\ref{fig:band} are available from Ref.~\cite{Sykes1973}.

\vspace{0.2cm}
\begin{acknowledgements}
	RL is supported by National Natural Science Foundation of China under Grant No.~12422503. JT is supported by National Natural Science Foundation of China under Grant No.~12405085 and by the Natural Science Foundation of Shanghai (Grant No.~24ZR1419300). JT would like to thank Ying Zhang for her love and support. The authors would like to disclose that GPT and Claude are used to assist to check the validity of certain mathematical derivations, and to polish the writing grammatically. The authors are fully responsible for the content of the manuscript. 
\end{acknowledgements}

\bibliographystyle{apsrev4-2}
\bibliography{ising_references}

%\clearpage
\onecolumngrid
\setcounter{secnumdepth}{2}
\setcounter{section}{0}
\setcounter{equation}{0}
\renewcommand{\thesection}{S\arabic{section}}
\renewcommand{\theequation}{S\arabic{equation}}
\setcounter{figure}{0}
\renewcommand{\thefigure}{S\arabic{figure}}
\begin{center}
\textbf{Supplemental Material}
\end{center}

\section{Droplet expansion}
\label{app:series}
In this SM we set up the Mayer expansion of the free energy in droplets, bound the size and the area of the Mayer clusters at each order of $u$, and show that the low-temperature series converges near $u=0$ to the physical free energy. We end with the pressure on strips of fixed width.

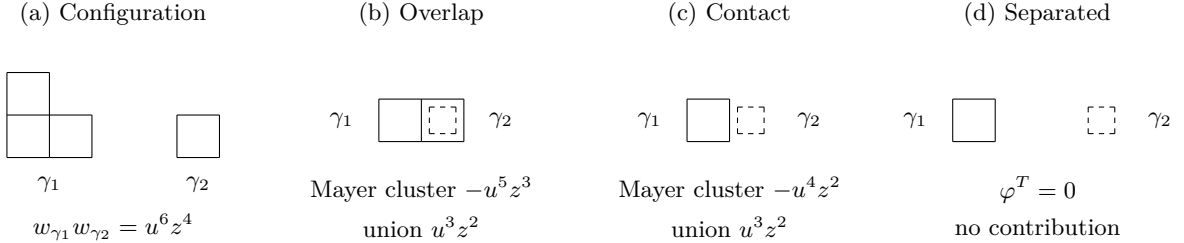
\begin{figure}[!h]
\centering
\setlength{\unitlength}{1pt}
\begin{picture}(458,98)
 \put(0,84){\makebox(110,12){\small (a) Configuration}}
 \put(116,84){\makebox(110,12){\small (b) Overlap}}
 \put(232,84){\makebox(110,12){\small (c) Contact}}
 \put(348,84){\makebox(110,12){\small (d) Separated}}
 \put(15,36){\line(1,0){16}}
 \put(15,52){\line(1,0){16}}
 \put(15,68){\line(1,0){16}}
 \put(31,36){\line(1,0){16}}
 \put(31,52){\line(1,0){16}}
 \put(15,36){\line(0,1){16}}
 \put(15,52){\line(0,1){16}}
 \put(31,36){\line(0,1){16}}
 \put(31,52){\line(0,1){16}}
 \put(47,36){\line(0,1){16}}
 \put(79,36){\line(1,0){16}}
 \put(79,52){\line(1,0){16}}
 \put(79,36){\line(0,1){16}}
 \put(95,36){\line(0,1){16}}
 \put(14,20){\makebox(34,12){\small $\gamma_1$}}
 \put(70,20){\makebox(34,12){\small $\gamma_2$}}
 \put(0,4){\makebox(110,12){\small $w_{\gamma_1}w_{\gamma_2}=u^6z^4$}}
 \put(155,42){\line(1,0){16}}
 \put(155,58){\line(1,0){16}}
 \put(171,42){\line(1,0){16}}
 \put(171,58){\line(1,0){16}}
 \put(155,42){\line(0,1){16}}
 \put(171,42){\line(0,1){16}}
 \put(187,42){\line(0,1){16}}
 \put(174,45){\dashbox{2}(10,10){}}
 \put(131,44){\makebox(20,12){\small $\gamma_1$}}
 \put(191,44){\makebox(20,12){\small $\gamma_2$}}
 \put(116,18){\makebox(110,12){\small Mayer cluster $-u^5z^3$}}
 \put(116,4){\makebox(110,12){\small union $u^3z^2$}}
 \put(271,42){\line(1,0){16}}
 \put(271,58){\line(1,0){16}}
 \put(271,42){\line(0,1){16}}
 \put(287,42){\line(0,1){16}}
 \put(290,45){\dashbox{2}(10,10){}}
 \put(247,44){\makebox(20,12){\small $\gamma_1$}}
 \put(307,44){\makebox(20,12){\small $\gamma_2$}}
 \put(232,18){\makebox(110,12){\small Mayer cluster $-u^4z^2$}}
 \put(232,4){\makebox(110,12){\small union $u^3z^2$}}
 \put(371,42){\line(1,0){16}}
 \put(371,58){\line(1,0){16}}
 \put(371,42){\line(0,1){16}}
 \put(387,42){\line(0,1){16}}
 \put(422,45){\dashbox{2}(10,10){}}
 \put(347,44){\makebox(20,12){\small $\gamma_1$}}
 \put(439,44){\makebox(20,12){\small $\gamma_2$}}
 \put(348,18){\makebox(110,12){\small $\varphi^T=0$}}
 \put(348,4){\makebox(110,12){\small no contribution}}
\end{picture}
\caption{Droplets and Mayer clusters. Each cell is a reversed spin. (a) A configuration whose reversed spins form two droplets, $\gamma_1$ and $\gamma_2$, which are its maximal connected pieces. Droplets of one configuration never share a site or touch, and the weight of the configuration is the product $w_{\gamma_1}w_{\gamma_2}$. (b)--(d) Pairs of droplets in the Mayer expansion~\eqref{eq:polymer}, with $\gamma_1$ solid and $\gamma_2$ dashed. In (b) the domino $\gamma_1$ and the single spin $\gamma_2$ share a site, and in (c) two single spins occupy neighboring sites, so each pair is a Mayer cluster with term $-w_{\gamma_1}w_{\gamma_2}$. In both cases the union of the two droplets is a domino with weight $u^3z^2$, which differs from the Mayer-cluster term because the Mayer cluster counts the boundary and the area of each droplet separately. A droplet repeated twice is also a Mayer cluster, with term $-\tfrac12w_\gamma^2$. In (d) the two droplets are compatible, $\varphi^T$ vanishes, and the pair contributes nothing to the free energy.}
\label{fig:clusters}
\end{figure}

We decompose a set of reversed spins into its maximal nearest-neighbor connected components $\gamma$, the droplets, with weights $w_\gamma=u^{m(\gamma)}z^{|\gamma|}$. Two droplets are compatible when they are disjoint and no site of one is a nearest neighbor of a site of the other, and we set $f(\gamma,\gamma')=-1$ for incompatible droplets, including $\gamma'=\gamma$, and $f(\gamma,\gamma')=0$ otherwise. For the $a\times b$ box $\Lambda_{a,b}$ with plus boundary spins, $P_{a,b}$ of~\eqref{eq:finiteP} is the partition function of a hard-core gas of droplets~\cite{GruberKunz1971}, and its logarithm is the Mayer expansion~\cite{Mayer1937,Brydges1986,Bissacot2010}
\begin{equation}
	\begin{aligned}
		\log P_{a,b}&=\sum_{k\ge1}\frac1{k!}\sum_{\gamma_1,\ldots,\gamma_k\subseteq\Lambda_{a,b}}\varphi^T(\gamma_1,\ldots,\gamma_k)\prod_{i=1}^k w_{\gamma_i},\\
		\varphi^T(\gamma_1,\ldots,\gamma_k)&=\sum_{G}\ \prod_{\{i,j\}\in G}f(\gamma_i,\gamma_j),
	\end{aligned}
	\label{eq:polymer}
\end{equation}
where $G$ runs over the connected graphs on the $k$ labels, and repeated droplets are allowed. A Mayer cluster is a tuple of droplets with connected incompatibility graph, namely with $\varphi^T\ne0$. The translates of a Mayer cluster form its translation class, so rotated or reflected Mayer clusters lie in different classes. Figure~\ref{fig:clusters} illustrates the difference between the droplets of a configuration and Mayer clusters.

At total half-perimeter $m$ a Mayer cluster has $k\le m/2$ droplets, each of bounded area. The union $S$ of its droplets is connected, so $S$ occupies every row and every column of its bounding box, of size $w_1\times w_2$, and each of these rows and columns contains at least two boundary bonds of $S$. Every boundary bond of $S$ is a boundary bond of one of the droplets, so
\begin{equation}
	2(w_1+w_2)\le|\partial S|\le\sum_j|\partial\gamma_j|=2m.
\end{equation}
Hence $w_1+w_2\le m$. Placing the lower-left corner of the bounding box at the origin selects one Mayer cluster in each translation class, and each of its at most $m/2$ droplets is then one of the finitely many subsets of a box with $w_1+w_2\le m$, so there are finitely many translation classes of Mayer clusters at order $u^m$. Since $w_1,w_2\le m$, every class fits in $\Lambda_{a,b}$ once $a,b\ge m$. A translation $\gamma_j\mapsto\gamma_j+t$ by a common $t\in\Z^2$ that keeps all droplets inside $\Lambda_{a,b}$ preserves their areas, their half-perimeters, which include the bonds to the fixed boundary spins, and their incompatibilities. So all members of a translation class $\mathcal C$ have the same term in~\eqref{eq:polymer}, with coefficient of $u^m$
\begin{equation}
	c_{\mathcal C}(z)=\frac1{k!}\,\varphi^T(\gamma_1,\ldots,\gamma_k)\,z^{\sum_j|\gamma_j|}\in\Q[z].
\end{equation}
If the bounding box of $\mathcal C$ is $w_1(\mathcal C)\times w_2(\mathcal C)$, its lower-left corner can sit at $a-w_1(\mathcal C)+1$ horizontal and $b-w_2(\mathcal C)+1$ vertical positions in $\Lambda_{a,b}$, so for $a,b\ge m$
\begin{equation}
	\begin{aligned}
		[u^m]\log P_{a,b}&=\sum_{\mathcal C}c_{\mathcal C}(z)\bigl(a-w_1(\mathcal C)+1\bigr)\bigl(b-w_2(\mathcal C)+1\bigr)\\
		&=ab\sum_{\mathcal C}c_{\mathcal C}(z)+\lambda_ma+\lambda'_mb+\kappa_m,
	\end{aligned}
\end{equation}
with $\lambda_m=-\sum_{\mathcal C}c_{\mathcal C}\,(w_2(\mathcal C)-1)$, $\lambda'_m=-\sum_{\mathcal C}c_{\mathcal C}\,(w_1(\mathcal C)-1)$ and $\kappa_m=\sum_{\mathcal C}c_{\mathcal C}\,(w_1(\mathcal C)-1)(w_2(\mathcal C)-1)$. Only the bulk term survives in $\Phi=\lim_{a,b\to\infty}(ab)^{-1}\log P_{a,b}$, so the coefficient $L_m$ in~\eqref{eq:series} is $L_m=\sum_{\mathcal C}c_{\mathcal C}(z)$. Each placement count obeys $\Delta_a\Delta_b\bigl[(a-w_1+1)(b-w_2+1)\bigr]=1$, and the four boxes in $\Delta_a\Delta_b\log P_{a,b}$ have sides at least $m$ when $a,b\ge m+1$, which gives~\eqref{eq:bulkdiff}.

The largest total area of $k$ droplets with total half-perimeter $m$ is
\begin{equation}
	\max_{\substack{m_1+\cdots+m_k=m\\m_i\ge2}}\sum_i A_{m_i}=A_{m-2(k-1)}+(k-1).
	\label{eq:kfactors}
\end{equation}
To see this, we move one unit of half-perimeter from a smaller factor with $m_i>2$ to a larger factor. Since $A_{t+1}-A_t=\lfloor(t+1)/2\rfloor$ is nondecreasing in $t$, the total area can only grow. For $k\ge2$ the right-hand side is largest at $k=2$, where it equals the bound $B_m$ of~\eqref{eq:gap}.

We next show that the low-temperature series is the Taylor expansion of the physical free energy at fixed $|z|<1$. We use an $N\times N$ torus and expand in the fugacity at fixed $u$,
\begin{equation}
	Z_N=e^{N^2(2K+H)}P_N(u,z),\qquad\frac{1}{N^2}\log P_N(u,z)=\sum_{n\ge1}b_{n,N}(u)z^n.
	\label{eq:highfieldfinite}
\end{equation}
At fixed $n$, $b_{n,N}$ is a polynomial in $u$ of degree at most $2n$, since a set $S$ of $n$ reversed spins has $m(S)=2n-e(S)\le2n$, and every Mayer cluster of total area $n$ obeys this bound too. The union of the droplets of such a Mayer cluster is a connected set of at most $n$ sites. For $N>n$ it therefore neither winds around the torus nor touches its periodic images, so its droplets, their half-perimeters, and their incompatibilities are the same as on $\Z^2$, and each translation class has exactly $N^2$ members, one for each position of the lower-left corner of its bounding box. By the Mayer expansion~\eqref{eq:polymer} on the torus, $b_{n,N}=b_n$ for $N>n$, with the $N$-independent polynomial
\begin{equation}
	b_n(u)=\sum_{\mathcal C}\frac1{k!}\,\varphi^T(\gamma_1,\ldots,\gamma_k)\,u^{\sum_jm(\gamma_j)},
\end{equation}
where $\mathcal C$ runs over the translation classes of Mayer clusters on $\Z^2$ with $\sum_j|\gamma_j|=n$, each represented by one tuple $(\gamma_1,\ldots,\gamma_k)$.

For real $u\in(0,1]$, the Lee--Yang circle theorem places the $N^2$ zeros $\zeta_j(u)$ of the polynomial $z\mapsto P_N(u,z)$ on the unit circle, by Theorem~3 in Sec.~IV of~\cite{LeeYang1952}, proved in its Appendix~II. The polynomial has constant term one and leading coefficient one, so $P_N(u,z)=\prod_j\bigl(1-z/\zeta_j(u)\bigr)$. Expanding $\log\bigl(1-z/\zeta_j(u)\bigr)=-\sum_nz^n/\bigl(n\,\zeta_j(u)^n\bigr)$, which converges for $|z|<1=|\zeta_j(u)|$, and using~\eqref{eq:highfieldfinite}, we have
\begin{equation}
	b_{n,N}(u)=-\frac{1}{nN^2}\sum_j\zeta_j(u)^{-n},\qquad |b_{n,N}(u)|\le\frac1n.
	\label{eq:LYbound}
\end{equation}
The zeros depend on $N$, but for $N>n$ this combination is the $N$-independent $b_n(u)$, so $|b_n(u)|\le1/n$. Every term of $b_n$ carries a factor $u^{\sum_jm(\gamma_j)}$, and each droplet has $m(\gamma_j)\ge2$, since $1\le|\gamma_j|\le A_{m(\gamma_j)}$ by~\eqref{eq:isoperimetry}. Hence $b_n(0)=0$, and the bound holds at $u=0$ too. By~\eqref{eq:highfieldfinite}, $N^{-2}\log P_N(u,z)-\sum_nb_n(u)z^n=\sum_{n\ge N}\bigl(b_{n,N}(u)-b_n(u)\bigr)z^n$, because $b_{n,N}=b_n$ for $n<N$. Both coefficients are at most $1/n$ in modulus, so the difference is at most $\sum_{n\ge N}2|z|^n/n\le2|z|^N/[N(1-|z|)]$, which tends to zero as $N\to\infty$ for $|z|<1$. Hence $\lim_{N\to\infty}N^{-2}\log P_N(u,z)=\sum_nb_n(u)z^n$ for real $u\in[0,1]$ and $|z|<1$.

To reach small complex $u$, we expand $b_n$ in Chebyshev polynomials, $b_n(u)=\sum_{j=0}^{2n}c_{nj}T_j(2u-1)$. By the cosine-integral formula for $c_{nj}$ and~\eqref{eq:LYbound}, $|c_{n0}|\le1/n$ and $|c_{nj}|\le2/n$ for $j\ge1$. We define
\begin{equation}
	R_\epsilon=\max_{|u|\le\epsilon}\ \max_{w+w^{-1}=2(2u-1)}|w|.
\end{equation}
The two roots $w$ are reciprocal, so $R_\epsilon\ge1$, and $R_\epsilon\to1$ as $\epsilon\to0$. Since $T_j((w+w^{-1})/2)=(w^j+w^{-j})/2$,
\begin{equation}
	|b_n(u)|\le(4+1/n)R_\epsilon^{2n}\le5R_\epsilon^{2n},\qquad |u|\le\epsilon.
\end{equation}
For any $\rho<1$ we choose $\epsilon>0$ with $\rho R_\epsilon^2<1$. Then $|b_n(u)z^n|\le5(\rho R_\epsilon^2)^n$ on $|u|\le\epsilon$, $|z|\le\rho$, so $\sum_nb_n(u)z^n$ converges uniformly there by the Weierstrass M-test. Its terms are polynomials in $u$ and $z$, so the sum is analytic in $(u,z)$ jointly, and it equals the physical $\Phi$ on the real ferromagnetic part of that domain. Its Taylor coefficients in $u$ are $L_m(z)=\lim_{a,b\to\infty}(ab)^{-1}[u^m]\log P_{a,b}$, because at each bidegree the torus and the box with plus boundary spins count the same finite Mayer clusters. By the area bound~\eqref{eq:isoperimetry}, $[u^m]b_n=0$ for $n>A_m$, so each $L_m$ is a polynomial. As a concrete example, the polynomials $L_2,\ldots,L_{11}$ are listed as $\psi_2(\mu),\ldots,\psi_{11}(\mu)$, with $\mu=z$, for the simple quadratic lattice in the Appendix on p.~1074 of~\cite{Sykes1973}.

Finally we consider a strip of width $w$ and length $\ell$, with periodic boundary conditions in both directions. Its partition function is the trace of the $\ell$th power of a $2^w\times2^w$ transfer matrix $T_w$. This is the construction of Eqs.~(2.1.6)--(2.1.10) on p.~33 of~\cite{Baxter1982}, with a single spin replaced by a row of $w$ spins, as in Eqs.~(7.2.1)--(7.2.3) on p.~90 for the square lattice in zero field. In a field, each row configuration $(\sigma_1,\ldots,\sigma_w)$ carries the extra factor $e^{H\sum_i\sigma_i}$, which multiplies the zero-field $T_w$ by a diagonal matrix, so the entries of $T_w$ are products of $e^{\pm K}$ and $e^{\pm H}$. As $\ell\to\infty$ at fixed $w$, the argument of Eq.~(2.1.16) on p.~34 and Eq.~(7.2.5) on p.~91 of~\cite{Baxter1982} gives $p_w=w^{-1}\log\lambda_w$, where $\lambda_w$ is the largest eigenvalue of $T_w$. It is a root of the characteristic polynomial of $T_w$, whose coefficients are polynomials in $e^{\pm K}$ and $e^{\pm H}$, so $\lambda_w$ is algebraic over $\Q(e^K,e^H)$. Since algebraic functions, $e^K$, and the logarithm are differentially algebraic, $p_w$ is differentially algebraic in $K$ by Proposition~3.7 of~\cite{Manssour2024}.

\section{Rational, algebraic, and transcendental fugacities}
\label{app:arithmetic}
In this SM we recall the known bounds on the denominators of power series that obey algebraic differential equations. We then give the details of the argument at algebraic and transcendental fugacities, extend it to the magnetization, and explain why fugacities that are algebraic integers escape our arguments.

Let $F$ be a number field and $g(u)=\sum_{n\ge0}g_nu^n\in F[[u]]$. For a nonzero algebraic number $g_n$, the denominator $d_n$ is the least positive integer such that $d_ng_n$ is an algebraic integer, and $d_n=1$ for $g_n=0$. If $g$ satisfies a nonzero algebraic differential equation with complex coefficients, it also satisfies one with coefficients in $F$, by statement~(6) in Sec.~25 of~\cite{Mahler1976}, see also Proposition~8 of Chap.~II of~\cite{Kolchin1973} and its Corollary. Its denominators then obey~\eqref{eq:polya} by the P\'olya--Popken theorem, which is Theorem~18 of~\cite{Mahler1976}. The stronger Sibuya--Sperber theorem bounds the coefficients at every fixed finite place $v$,
\begin{equation}
	v(g_n)\ge-C_v(n+1),
	\label{eq:SS}
\end{equation}
with a constant $C_v$ independent of $n$, which depends only on $g$, its equation, and $v$~\cite{SibuyaSperber1981}. Both bounds are stated in Sec.~1 of~\cite{KrattenthalerRivoal2025}, the first as its Eq.~(1.2), where \eqref{eq:SS} is identified with Eq.~(III$'$) on p.~112 of~\cite{SibuyaSperber1981}.

A finite place $v$ of a number field $F$ comes from a nonzero prime ideal of the algebraic integers of $F$, which contains exactly one rational prime $\ell$. We normalize $v$ to extend $v_\ell$, so that $v(\ell)=1$, as in Chap.~I, Sec.~8, p.~44, and Chap.~II, Sec.~8, p.~165, of~\cite{Neukirch1999}. It obeys $v(xy)=v(x)+v(y)$ and $v(x+y)\ge\min\{v(x),v(y)\}$, with equality when $v(x)\ne v(y)$. Hence a finite sum in which one term $x_0$ has a smaller valuation than every other term has valuation $v(x_0)$. By the inequality, the other terms together have valuation above $v(x_0)$, and by the equality case, adding $x_0$ gives $v(x_0)$.

Let $\alpha\ne0$ be algebraic but not an algebraic integer, for instance a rational $A/B$ with $B>1$. Then some finite place $v$ of $F=\Q(\alpha)$ has $v(\alpha)=-a<0$, because the algebraic integers of $F$ are the elements with $v\ge0$ at every finite place, as shown in the proof of Proposition~(11.5) in Chap.~I, p.~68, of~\cite{Neukirch1999}. By the decomposition~\eqref{eq:split},
\begin{align}
	v\bigl(\alpha^{r^2}\bigr)&=-ar^2,\\
	v\bigl(S_{2r}(\alpha)-\alpha^{r^2}\bigr)&\ge-a(r^2-1),\\
	v\Bigl(\frac{R_{2r}(\alpha)}{2r}\Bigr)&\ge-a(r^2-2r+2)-v_\ell(2r).
\end{align}
For sufficiently large $r$, $a(2r-2)>v_\ell(2r)$, so $\alpha^{r^2}$ is the only term with the smallest valuation, and
\begin{equation}
	v\bigl(L_{2r}(\alpha)\bigr)=-ar^2.
	\label{eq:algebraicvaluation}
\end{equation}
For $\alpha=A/B$ and $v=v_\ell$ with $\ell\mid B$ this is~\eqref{eq:valuation}. Since $d_{2r}L_{2r}(\alpha)$ is an algebraic integer, $v_\ell(d_{2r})=v(d_{2r})\ge ar^2$, so $\log d_{2r}\ge ar^2\log\ell$ contradicts~\eqref{eq:polya}. Equation~\eqref{eq:algebraicvaluation} also contradicts~\eqref{eq:SS} directly. Hence $\Phi(u,\alpha)$ is differentially transcendental in $u$.

The magnetization~\eqref{eq:observables} needs a separate argument, because it involves a derivative with respect to the parameter $z$ taken before $z$ is fixed. Its coefficient of $u^{2r}$ is $-2\alpha L_{2r}'(\alpha)$. On the infinite subsequence $\ell\nmid r$, the leading coefficient of $\alpha L_{2r}'(\alpha)$ is the $v$-unit $r^2$. The operator $z\partial_z$ keeps the coefficients integral and the degree unchanged, so the comparison that led to \eqref{eq:algebraicvaluation} now yields
\begin{equation}
 v\bigl([u^{2r}](M-1)\bigr)=v(2)-ar^2,
 \qquad \ell\nmid r,
 \label{eq:Mvaluation}
\end{equation}
for sufficiently large $r$. This contradicts \eqref{eq:SS}, and for rational $\alpha$ it also contradicts \eqref{eq:polya} directly.

Let $\beta\in(0,1)$ be transcendental. An equation~\eqref{eq:DA} for $\Phi(u,\beta)$ is a complex linear relation $\sum_ic_i\mathcal M_i=0$, with coefficients $c_i$ not all zero, among its monomials $\mathcal M_i$, $i=1,\ldots,N$, evaluated at $z=\beta$. Since every $L_m$ lies in $\Q[z]$, every coefficient $t_{ni}$ in $\mathcal M_i=\sum_{n\ge0}t_{ni}(z)\,u^n$ lies in $\Q[z]$, and~\eqref{eq:linear} reads, with one row for each order $u^n$,
\begin{equation}
	\begin{pmatrix}
		t_{01}(z)&t_{02}(z)&\cdots&t_{0N}(z)\\
		t_{11}(z)&t_{12}(z)&\cdots&t_{1N}(z)\\
		\vdots&\vdots&&\vdots
	\end{pmatrix}
	\begin{pmatrix}c_1\\c_2\\\vdots\\c_N\end{pmatrix}=0.
	\label{eq:matrix}
\end{equation}
At any rational $z_0\in(0,1)$ the $\mathcal M_i$ are linearly independent over $\C$, because $\Phi(u,z_0)$ is differentially transcendental, so at $z=z_0$ the only solution of~\eqref{eq:matrix} is $c=0$. The rows of the matrix then span $\C^N$, so some $N$ of them, $n_1<\cdots<n_N$, are linearly independent, and the minor
\begin{equation}
	D(z)=\det\bigl(t_{n_ki}(z)\bigr)_{k,i=1}^{N}\in\Q[z]
	\label{eq:minor}
\end{equation}
obeys $D(z_0)\ne0$. Thus $D$ is a nonzero polynomial with rational coefficients, which does not vanish at the transcendental $\beta$, so at $z=\beta$ the rows $n_1,\ldots,n_N$ of~\eqref{eq:matrix} force $c_i=0$. Since the $c_i$ are not all zero, $\Phi(u,\beta)$ fails to obey~\eqref{eq:DA}. Combining this with the algebraic case, we find that $\Phi(u,z)$ is differentially transcendental in $u$ at every $z\in(0,1)$ that is not an algebraic integer. The same argument applies to $M$, whose coefficients in $u$ also lie in $\Q[z]$.

Both arguments fail at an algebraic integer $\alpha$, such as $\alpha=(\sqrt5-1)/2\in(0,1)$, a root of $x^2+x-1$. Since $\alpha$ is an algebraic integer, $v(\alpha)\ge0$ at every finite place $v$ of $\Q(\alpha)$, by the proof of Proposition~(11.5) in Chap.~I, p.~68, of~\cite{Neukirch1999}. Hence the square term $\alpha^{r^2}$ carries no denominator, and by~\eqref{eq:integrality} $2rL_{2r}(\alpha)$ is an algebraic integer, so the denominators divide $2r$ and are consistent with~\eqref{eq:polya}. The minor $D(z)$ can vanish at an algebraic $\alpha$, so the transcendental argument gives nothing either. Algebraic integers are roots of monic integer polynomials, of which there are countably many, so these fugacities, and the fields $H=-\tfrac12\log\alpha$ they define, form a countable set. At these fields we leave the differential algebraicity of the free energy undecided.

\section{Exact solutions}
\label{app:exact}
In this SM we show that an algebraic differential equation for the pressure on a short interval of couplings would hold at all couplings, which is excluded by our result. We also show that solutions of Onsager's type and closed forms built from elliptic, hypergeometric, or Painlev\'e functions obey such equations, as does the pressure of mean-field theory.

We fix $H>0$, and with it $z$, and suppose that the pressure obeys
\begin{equation}
	\mathcal P\bigl(K,p,\partial_Kp,\ldots,\partial_K^sp\bigr)=\sum_{\mu,\{\nu_j\}}c_{\mu,\{\nu_j\}}\,K^{\mu}\prod_{j=0}^{s}(\partial_K^jp)^{\nu_j}=0
	\label{eq:intervalODE}
\end{equation}
for $K$ in an interval $(K_1,K_2)$, with a finite sum as in~\eqref{eq:DA} and complex coefficients $c_{\mu,\{\nu_j\}}$, not all zero. By Corollary~1.4 of~\cite{Ott2020}, whose $\beta$ and $h$ are our $K$ and $H$, the pressure is analytic in a neighborhood of every point $(K,H)$ with $H\ne0$. At our fixed $H$, the function $K\mapsto p(K,H)$ is therefore real analytic at every point of the half-line $K>0$. The left-hand side $F(K)$ of~\eqref{eq:intervalODE} is a polynomial in $K$ and in derivatives of $p$, so $F$ is real analytic on $K>0$ as well. It vanishes on $(K_1,K_2)$, and the half-line is connected, so by the identity theorem for real-analytic functions $F(K)=0$ for every $K>0$. The equation therefore holds at arbitrarily large finite $K$, that is, at arbitrarily small $u=e^{-4K}>0$, where $p=2K+H+\Phi(u,z)$ and $\Phi$ is given by the series~\eqref{eq:series}, which converges near $u=0$ by Sec.~\ref{app:series}. Since $K=-\tfrac14\log u$ is differentially algebraic in $u$ and the class is closed under arithmetic operations and composition by Propositions~3.6 and~3.7 of~\cite{Manssour2024}, $\Phi(u,z)=p+\tfrac12\log u-H$ obeys an equation~\eqref{eq:DA} for $0<u<1$. Its left-hand side is analytic near $u=0$ and vanishes for small $u>0$, so it vanishes identically there, and the series~\eqref{eq:series} itself obeys~\eqref{eq:DA}. By Sec.~\ref{app:arithmetic}, this is impossible at every fugacity that is not an algebraic integer.

We now show that pressures of Onsager's type~\eqref{eq:FF}, with finitely many terms, constants $c_j$, trigonometric polynomials $Q_j$ in $\theta_1,\ldots,\theta_k$, and $R$ and the coefficients of the $Q_j$ algebraic in $e^{\pm K}$ and $e^{\pm H}$, obey algebraic differential equations. Onsager's solution has this form, with $R=\log2$, $k=2$, and a single term with $c_1=\tfrac12$ and $Q_1=\cosh^22K-\sinh2K\,(\cos\theta_1+\cos\theta_2)$~\cite{Onsager1944}, and so does every translation-invariant model of free fermions with finite-range couplings, with $Q_j$ the determinant of its coupling matrix in momentum space~\cite{Kasteleyn1961,FanWu1970}. We fix $z$ and take $u$ in an interval on which no $Q_j$ vanishes on the torus $|e^{i\theta_a}|=1$, as for gapped free fermions. For Onsager's solution, $Q_1\ge(\sinh2K-1)^2$ on the torus, so this excludes only the critical point $\sinh2K=1$, Eq.~(6.2.17) on p.~78 of~\cite{Baxter1982}. Each $\log Q_j$ then has a continuous branch on $[-\pi,\pi]^k$ times that interval, unique up to a constant in $2\pi i\Z$ that drops out of the $u$-derivative, and differentiating under the integral gives
\begin{equation}
 \partial_u\int\frac{\dd^k\theta}{(2\pi)^k}\,\log Q_j
 =\int\frac{\dd^k\theta}{(2\pi)^k}\,\frac{\partial_uQ_j}{Q_j}.
 \label{eq:period}
\end{equation}
With $x_a=e^{i\theta_a}$, the right-hand side is the constant term in $x_1,\ldots,x_k$ of the Laurent expansion, in the region that contains the torus, of the rational function $\partial_uQ_j/Q_j$, whose coefficients are algebraic in $u$ because $e^{\pm K}=u^{\mp1/4}$ and $e^{\pm H}=z^{\mp1/2}$ are. Such constant terms obey linear differential equations with polynomial coefficients, the Picard--Fuchs equations of~\eqref{eq:period}~\cite{Lipshitz1988,Zeilberger1990}. Antiderivatives, sums, and products of such functions, and algebraic functions, obey such equations too, so~\eqref{eq:FF} obeys a linear differential equation in $u$ on that interval, and hence an algebraic differential equation in $K$. Since such an equation on an interval of couplings extends to all couplings, a pressure of the form~\eqref{eq:FF} can match the Ising pressure on an interval of $K$ only at fugacities that are algebraic integers. The same holds for closed forms built through arithmetic operations and composition from elliptic, hypergeometric, or Painlev\'e functions with fixed parameters and with arguments, including the modulus or nome of elliptic functions, that are differentially algebraic in $K$ at fixed $H$, by Propositions~3.6 and~3.7 of~\cite{Manssour2024}. Parameters that vary with $K$ are not covered, since for example Gauss's theorem $_2F_1(a,b;c;1)=\Gamma(c)\Gamma(c-a-b)/[\Gamma(c-a)\Gamma(c-b)]$ brings in the Gamma function, which is differentially transcendental~\cite{Holder1886}.

Finally we treat mean-field theory. The Curie--Weiss model is the mean-field Hamiltonian, Eq.~(3.1.3) on p.~39 of~\cite{Baxter1982}, with $qJ$ there equal to our $J$. Its magnetization obeys $\bar M=\tanh(K\bar M+H)$, Eq.~(3.1.18) on p.~41, and its pressure is $\bar p=\tfrac12\log[4/(1-\bar M^2)]-\tfrac12K\bar M^2$, Eq.~(3.1.21) on p.~42. After differentiating, we obtain $\dd \bar M/\dd K=\bar M(1-\bar M^2)/[1-K(1-\bar M^2)]$ and $\partial_K\bar p=\bar M^2/2$. The first equation makes $\bar M$ differentially algebraic, and $\bar p$, an antiderivative of $\bar M^2/2$, is differentially algebraic as well.

More generally, suppose that finitely many order parameters $y_1,\ldots,y_n$ are fixed by equations $F_i(K,y_1,\ldots,y_n)=0$, $i=1,\ldots,n$, and that the mean-field pressure is $\bar p=G(K,y_1,\ldots,y_n)$, where the $F_i$ and $G$ are elementary functions, built from their arguments and constants by arithmetic operations, exponentials, logarithms, and algebraic functions. For the Curie--Weiss model, $n=1$, $F_1=y_1-\tanh(Ky_1+H)$, and $G=\tfrac12\log[4/(1-y_1^2)]-\tfrac12Ky_1^2$. Let $\mathcal F$ be the field generated over $\C(K)$ by the $y_j$ and by the finitely many exponentials, logarithms, and algebraic functions that occur in the $F_i$ and $G$. On a nondegenerate branch, where $\det(\partial F_i/\partial y_j)\ne0$, implicit differentiation gives every $\dd y_j/\dd K$ as an element of $\mathcal F$, and the chain rule does the same for the derivative of every adjoined function. Hence $\mathcal F$ is closed under $\dd/\dd K$ and has finite transcendence degree $t$ over $\C(K)$, so $\bar p,\partial_K\bar p,\ldots,\partial_K^t\bar p$ obey a nonzero polynomial relation over $\C(K)$, and $\bar p$ is differentially algebraic. At every field outside the countable set the lattice pressure is not differentially algebraic, so no mean-field approximation of this kind can reproduce it on an interval of $K$ there.

\section{Every dimension \texorpdfstring{$d\ge2$}{d>=2}}
\label{app:dims}
In this SM we carry the argument over to the hypercubic lattice $\Z^d$, with cubes in place of squares and the isoperimetric inequality~\eqref{eq:LW} in place of the area gap.

On $\Z^d$ the aligned state has pressure $dK+H$, and we write $p=dK+H+\Phi(u,z)$ with $\Phi=\sum_mL_m(z)u^m$. A set $S$ of reversed spins contributes $u^{m(S)}z^{|S|}$ with $m(S)=|\partial S|/2=d|S|-e(S)$, where $e(S)$ counts the bonds inside $S$, so every nonempty set has $m(S)\ge d$.

By the Loomis--Whitney inequality, Theorem~2 on p.~961 of~\cite{LoomisWhitney1949}, and the inequality between arithmetic and geometric means, \eqref{eq:LW} follows, as stated in Theorem~6 on p.~6 of~\cite{EllisFriedgutKindlerYehudayoff2016}. Equality holds only for a cube, by Theorem~8 on p.~7 of~\cite{EllisFriedgutKindlerYehudayoff2016} with $\varepsilon=0$. Therefore, at half-perimeter $m=dr^{d-1}$ the largest volume is $r^d$ which is reached by only the cube of side $r$. For every set with half-perimeter $m$, connected or not,~\eqref{eq:LW} reads $|S|\le g(m)$ with $g(x)=(x/d)^{d/(d-1)}$.

A Mayer cluster of $k\ge2$ droplets with half-perimeters $m_j\ge d$ and $\sum_jm_j=m$ has total volume at most $\sum_jg(m_j)$. The function $g$ is convex with $g(d)=1$, and $g(x)-g(x-d)\ge g(2d)-g(d)=2^{d/(d-1)}-1\ge1$ for $x\ge2d$. The sum is therefore largest when all droplets but one are single spins, and it is at most $g(m-d)+1$. At $m=dr^{d-1}$, by the mean-value theorem,
\begin{equation}
 r^d-g(m-d)=r^d-\bigl(r^{d-1}-1\bigr)^{d/(d-1)}
 \ge\frac{d}{d-1}\bigl(r^{d-1}-1\bigr)^{1/(d-1)}\ge\frac{d}{d-1}(r-1),
 \label{eq:dgap}
\end{equation}
so every such Mayer cluster has total volume at most $r^d-\tfrac{d}{d-1}(r-1)+1$. In two dimensions this bound is weaker by one than \eqref{eq:gap}.

For a box $\Lambda$ of $a_1\times\cdots\times a_d$ sites with plus boundary spins we set $P=\sum_{S\subseteq\Lambda}u^{m(S)}z^{|S|}$, as in~\eqref{eq:finiteP}. Every droplet costs at least $u^d$, so a Mayer cluster at order $u^m$ has at most $m/d$ droplets. Let $w_1\times\cdots\times w_d$ be the size of the bounding box of the union $S$ of its droplets. Since $S$ is connected, it contains a site at each of the $w_b$ values of the $b$th coordinate in its bounding box. The rows of sites through these $w_b$ sites parallel to any other axis are distinct, and each contains at least two boundary bonds of $S$. Every boundary bond of $S$ is a boundary bond of one of the droplets, so $2w_b\le|\partial S|\le2m$ and $w_b\le m$. Placing a corner of the bounding box at the origin selects one Mayer cluster in each translation class, and each of its droplets is then one of the finitely many subsets of a cube of side $m$, so there are finitely many translation classes of Mayer clusters at order $u^m$. Since $w_b\le m$, every class fits in $\Lambda$ once all $a_b\ge m$. The logarithmic derivative $u\partial_u\log P$ has integer coefficients in every dimension, so $m[u^m]\log P\in\Z[z]$. Above the Mayer-cluster bound only single droplets contribute, each with coefficient one per translate, and at order $u^{dr^{d-1}}$ the coefficient of $z^{r^d}$ counts translates of the cube. In $\Lambda$, a translation class of Mayer clusters with bounding box $w_1\times\cdots\times w_d$ fits in $\prod_b(a_b-w_b+1)$ positions, and since $\Delta_{a_1}\cdots\Delta_{a_d}\prod_b(a_b-w_b+1)=1$, the mixed difference $\Delta_{a_1}\cdots\Delta_{a_d}\log P$ at all $a_b\ge m+1$ extracts $L_m$ as an integer combination of box coefficients, exactly as in Sec.~\ref{app:series}. Hence $mL_m\in\Z[z]$, the coefficients of $L_m$ above the Mayer-cluster bound are integers, and $[z^{r^d}]L_{dr^{d-1}}=1$.

For $z=A/B$ and a prime $\ell$ with $a_\ell=v_\ell(B)>0$, the cube term has $v_\ell=-a_\ell r^d$, the other terms above the Mayer-cluster bound have $v_\ell\ge-a_\ell(r^d-1)$, and the terms below it obey
\begin{equation}
 v_\ell\ge-a_\ell\Bigl[r^d-\tfrac{d}{d-1}(r-1)+1\Bigr]-v_\ell\bigl(dr^{d-1}\bigr).
\end{equation}
For large $r$ the cube term is the only term with the most negative valuation, so by the property of such sums in Sec.~\ref{app:arithmetic}
\begin{equation}
 v_\ell\bigl(L_{dr^{d-1}}(A/B)\bigr)=-a_\ell r^d=-a_\ell\Bigl(\frac md\Bigr)^{d/(d-1)}.
 \label{eq:dvaluation}
\end{equation}
This valuation decreases faster than linearly in the order $m$, which violates~\eqref{eq:SS}, and the corresponding denominators violate \eqref{eq:polya} as well.

The other ingredients of the two-dimensional argument carry over with the cube subsequence $m=dr^{d-1}$. Convergence of the low-temperature series near $u=0$ is established with the Lee--Yang theorem as in Sec.~\ref{app:series}. On the torus of $N^d$ sites we write $Z_N=e^{N^d(dK+H)}P_N(u,z)$ and $N^{-d}\log P_N=\sum_{n\ge1}b_{n,N}(u)z^n$. The union of the droplets of a Mayer cluster of total volume $n$ is again a connected set of at most $n$ sites, so for $N>n$ each translation class has exactly $N^d$ members, one for each position of the corner of its bounding box with the smallest coordinates, and $b_{n,N}=b_n$, where $b_n$ is the sum over translation classes on $\Z^d$. The Lee--Yang theorem holds on every ferromagnetic graph, by the general theorem of Appendix~II of~\cite{LeeYang1952}, so for real $u\in(0,1]$ it places the $N^d$ zeros $\zeta_j(u)$ of the polynomial $z\mapsto P_N(u,z)$ on the unit circle. The polynomial again has constant term one and leading coefficient one, so $P_N(u,z)=\prod_j\bigl(1-z/\zeta_j(u)\bigr)$. Expanding $\log\bigl(1-z/\zeta_j(u)\bigr)=-\sum_nz^n/\bigl(n\,\zeta_j(u)^n\bigr)$, which converges for $|z|<1=|\zeta_j(u)|$, we obtain $b_{n,N}(u)=-(nN^d)^{-1}\sum_j\zeta_j(u)^{-n}$ as in~\eqref{eq:LYbound}, so $|b_{n,N}(u)|\le1/n$, and $|b_n(u)|\le1/n$ since $b_{n,N}=b_n$ for $N>n$. At $u=0$ every droplet carries at least $u^d$, so $b_n(0)=0$. As in Sec.~\ref{app:series}, $N^{-d}\log P_N(u,z)-\sum_nb_n(u)z^n=\sum_{n\ge N}(b_{n,N}-b_n)z^n$ has modulus at most $\sum_{n\ge N}2|z|^n/n$, so $\lim_{N\to\infty}N^{-d}\log P_N(u,z)=\sum_nb_n(u)z^n$ for real $u\in[0,1]$ and $|z|<1$. The coefficients $b_n(u)$ now have degree at most $dn$, and the Chebyshev bound becomes $|b_n(u)|\le(2d+1)R_\epsilon^{dn}$ for $|u|\le\epsilon$. For any $\rho<1$ we choose $\epsilon>0$ with $\rho R_\epsilon^d<1$. Then $|b_n(u)z^n|\le(2d+1)(\rho R_\epsilon^d)^n$ on $|u|\le\epsilon$, $|z|\le\rho$, so $\sum_nb_n(u)z^n$ converges uniformly there by the Weierstrass M-test, and its sum is analytic in $(u,z)$ jointly. The arguments of Sec.~\ref{app:arithmetic} for algebraic and transcendental fugacities and those of Sec.~\ref{app:exact} for exact solutions apply verbatim, since the pressure is analytic in $K$ at fixed $H\ne0$ on $\Z^d$ as well, by Corollary~1.4 of~\cite{Ott2020}, which holds for every $d\ge1$. For a transcendental fugacity in particular, the entries of the matrix~\eqref{eq:matrix} again lie in $\Q[z]$, since $mL_m\in\Z[z]$ on $\Z^d$, and the rational case on $\Z^d$ gives a minor~\eqref{eq:minor} with $D(z_0)\ne0$ at a rational $z_0$. For the magnetization, the leading coefficient of $\alpha L'_{dr^{d-1}}(\alpha)$ is $r^d$, a $v$-unit when $\ell\nmid r$.

\end{document}